\documentclass[aps,prl,preprint,tightenlines,superscriptaddress,showpacs,byrevtex,subfigure]{revtex4}

\usepackage{graphicx}
\usepackage{color}

\begin{document}
\begin{flushright}
KEK Preprint 2009-20\\
Belle Preprint 2009-20\\
NTLP  Preprint 2009-02
\end{flushright}

\title{ \quad\\[0.5cm] 
Search for Lepton Flavor and
Lepton Number Violating $\tau$ 
{Decays}\\
into a Lepton and Two Charged  Mesons
}

\begin{abstract}
We search for lepton flavor
and lepton number violating $\tau$ decays 
into a lepton ($\ell$ = electron or muon) and two charged mesons 
($h,h' = \pi^\pm$ or $K^\pm$),
{$\tau^-\to\ell^-h^+h'^-$
and $\tau^-\to\ell^+h^-h'^-$,}
using
671 fb$^{-1}$ of data collected 
with the Belle detector at the 
KEKB asymmetric-energy $e^+e^-$ collider. 
{We obtain
90\% C.L. upper limits on the branching fractions
in the range 
$(4.4-8.8)\times 10^{-8}$ 
for  $\tau\to ehh'$,
and $(3.3-16)\times 10^{-8}$
for $\tau\to\mu hh'$ {processes.}
These results improve {upon}
previously published upper limits by factors between
1.6 to 8.8.}
\end{abstract}
\affiliation{Budker Institute of Nuclear Physics, Novosibirsk, Russian Federation}
\affiliation{Chiba University, Chiba, Japan}
\affiliation{University of Cincinnati, Cincinnati, OH, USA}
\affiliation{T. Ko\'{s}ciuszko Cracow University of Technology, Krakow, Poland}
\affiliation{The Graduate University for Advanced Studies, Hayama, Japan}
\affiliation{Hanyang University, Seoul, South Korea}
\affiliation{University of Hawaii, Honolulu, HI, USA}
\affiliation{High Energy Accelerator Research Organization (KEK), Tsukuba, Japan}
\affiliation{Hiroshima Institute of Technology, Hiroshima, Japan}
\affiliation{Institute of High Energy Physics, Chinese Academy of Sciences, Beijing, PR China}
\affiliation{Institute for High Energy Physics, Protvino, Russian Federation}
\affiliation{Institute of High Energy Physics, Vienna, Austria}
\affiliation{Institute for Theoretical and Experimental Physics, Moscow, Russian Federation}
\affiliation{J. Stefan Institute, Ljubljana, Slovenia}
\affiliation{Kanagawa University, Yokohama, Japan}
\affiliation{Institut f\"ur Experimentelle Kernphysik, Universit\"at Karlsruhe, Karlsruhe, Germany}
\affiliation{Korea University, Seoul, South Korea}
\affiliation{Kyungpook National University, Taegu, South Korea}
\affiliation{\'Ecole Polytechnique F\'ed\'erale de Lausanne, EPFL, Lausanne, Switzerland}
\affiliation{Faculty of Mathematics and Physics, University of Ljubljana, Ljubljana, Slovenia}
\affiliation{University of Maribor, Maribor, Slovenia}
\affiliation{Max-Planck-Institut f\"ur Physik, M\"unchen, Germany}
\affiliation{University of Melbourne, Victoria, Australia}
\affiliation{Nagoya University, Nagoya, Japan}
\affiliation{Nara Women's University, Nara, Japan}
\affiliation{National Central University, Chung-li, Taiwan}
\affiliation{National United University, Miao Li, Taiwan}
\affiliation{Department of Physics, National Taiwan University, Taipei, Taiwan}
\affiliation{H. Niewodniczanski Institute of Nuclear Physics, Krakow, Poland}
\affiliation{Nippon Dental University, Niigata, Japan}
\affiliation{Niigata University, Niigata, Japan}
\affiliation{University of Nova Gorica, Nova Gorica, Slovenia}
\affiliation{Novosibirsk State University, Novosibirsk, Russian Federation}
\affiliation{Osaka City University, Osaka, Japan}
\affiliation{Panjab University, Chandigarh, India}
\affiliation{University of Science and Technology of China, Hefei, PR China}
\affiliation{Seoul National University, Seoul, South Korea}
\affiliation{Shinshu University, Nagano, Japan}
\affiliation{Sungkyunkwan University, Suwon, South Korea}
\affiliation{School of Physics, University of Sydney, NSW 2006, Australia}
\affiliation{Excellence Cluster Universe, Technische Universit\"at M\"unchen, Garching, Germany}
\affiliation{Tohoku Gakuin University, Tagajo, Japan}
\affiliation{Tohoku University, Sendai, Japan}
\affiliation{Department of Physics, University of Tokyo, Tokyo, Japan}
\affiliation{Tokyo Metropolitan University, Tokyo, Japan}
\affiliation{Tokyo University of Agriculture and Technology, Tokyo, Japan}
\affiliation{IPNAS, Virginia Polytechnic Institute and State University, Blacksburg, VA, USA}
\affiliation{Yonsei University, Seoul, South Korea}
\author{Y.~Miyazaki} 
\affiliation{Nagoya University, Nagoya, Japan}
\author{H.~Aihara} 
\affiliation{Department of Physics, University of Tokyo, Tokyo, Japan}
\author{K.~Arinstein} 
\affiliation{Budker Institute of Nuclear Physics, Novosibirsk, Russian Federation}
\affiliation{Novosibirsk State University, Novosibirsk, Russian Federation}
\author{V.~Aulchenko} 
\affiliation{Budker Institute of Nuclear Physics, Novosibirsk, Russian Federation}
\affiliation{Novosibirsk State University, Novosibirsk, Russian Federation}
\author{T.~Aushev} 
\affiliation{\'Ecole Polytechnique F\'ed\'erale de Lausanne, EPFL, Lausanne, Switzerland}
\affiliation{Institute for Theoretical and Experimental Physics, Moscow, Russian Federation}
\author{A.~M.~Bakich} 
\affiliation{School of Physics, University of Sydney, NSW 2006, Australia}
\author{V.~Balagura} 
\affiliation{Institute for Theoretical and Experimental Physics, Moscow, Russian Federation}
\author{E.~Barberio} 
\affiliation{University of Melbourne, Victoria, Australia}
\author{A.~Bay} 
\affiliation{\'Ecole Polytechnique F\'ed\'erale de Lausanne, EPFL, Lausanne, Switzerland}
\author{K.~Belous} 
\affiliation{Institute for High Energy Physics, Protvino, Russian Federation}
% ...
\collaboration{The Belle Collaboration}
\noaffiliation

\maketitle
 \section{Introduction}

{Lepton flavor violation (LFV)
in charged lepton decays is forbidden or highly suppressed 
{even if} neutrino mixing is included.
However, LFV {appears} in {various} extensions of the Standard Model (SM),
such as supersymmetry, leptoquark and many other 
models{~\cite{cite:amon,cite:six_fremionic,cite:brignole,cite:susy1,cite:maria,cite:fukuyama,cite:Raidal,cite:maria2}.}
{Some of these models predict branching fractions
which, for
certain combinations of model parameters,
can be as high as $10^{-7}$;
{these rates are}
already accessible
in
high-statistics
{$B$-factory} experiments.
Here, we {search} for 
$\tau$ decays{~\footnotemark[2]}
into one lepton 
($\ell$ = electron or muon) and
two charged mesons 
($h, h' = \pi^\pm$ or $K^\pm$)
including lepton flavor 
and 
lepton number violation 
($\tau^-\to\ell^-h^-h'^+$  and $\tau^-\to\ell^+h^-h'^-$)
\footnotemark[3],
with a data sample of 671 fb$^{-1}$ collected at the $\Upsilon(4S)$ 
resonance
and 60 MeV below with the Belle detector at the KEKB 
asymmetric-energy   $e^+e^-$ collider~\cite{kekb}.
Previously, we
{reported}
90\% confidence level (C.L.) upper limits
{on} 
{these LFV}
branching fractions
using 158 fb${}^{-1}$ of data;
the results were
in the range (1.6$-$8.0)~$\times~10^{-7}$~\cite{lhh_belle}.
{The BaBar collaboration
has
also  obtained 90\% C.L.
upper limits
in the range 
{(0.7$-$4.8)~$\times~10^{-7}$}~\cite{lhh_babar}.}
using 221 fb${}^{-1}$ of data

\footnotetext[2]{{Throughout this paper,
charge-conjugate modes are {implied} unless stated  otherwise.}}

\footnotetext[3]{{The notation ``$\tau\to\ell hh'$ ``
indicates
both 
$\tau^-\to\ell^- h^+h'^-$ 
and
$\tau^-\to\ell^+ h^-h'^-$ modes.}}

The Belle detector is a large-solid-angle magnetic spectrometer that
consists of a silicon vertex detector (SVD),
a 50-layer central drift chamber (CDC),
an array of aerogel threshold Cherenkov counters (ACC),
a barrel-like arrangement of
time-of-flight scintillation counters (TOF),
and an electromagnetic calorimeter
comprised of CsI(Tl) {crystals (ECL), all located} inside
a superconducting solenoid coil
that provides a 1.5~T magnetic field.
An iron flux-return located {outside} the coil is instrumented to
detect $K_{\it{L}}^0$ mesons
and to identify muons (KLM).
The detector is described in detail elsewhere~\cite{Belle}.

Particle identification
is very important
{for this measurement.}
{We use 
hadron identification likelihood variables} 
based on
the ratio of the energy
deposited in the
ECL to the momentum measured in the SVD and CDC,
the shower shape in the ECL,
the particle range in the KLM,
the hit information from the ACC,
{the $dE/dx$ information in the CDC,}
and {the {particle} time-of-flight} from the TOF.
To distinguish hadron species,
we use likelihood ratios,
${\cal{P}}(i/j) = {\cal{L}}_i/({\cal{L}}_i + {\cal{L}}_{j})$,
where ${\cal{L}}_{i}$ (${\cal{L}}_{j}$)
is the likelihood for the detector response
to {a} track with flavor hypothesis $i$ ($j$).
{{For lepton identification,
we {form} likelihood ratios ${\cal P}(e)$~\cite{EID}
and ${\cal P}({\mu})$~\cite{MUID}
based on the}
electron and  muon probabilities, respectively,
{which are}
determined by
the responses of the appropriate subdetectors.}

In order to estimate the signal efficiency and 
optimize the event selection, 
we use {Monte Carlo (MC) simulated event samples}.
The signal and background events from generic $\tau^+\tau^-$ decays are 
 generated by KKMC/TAUOLA~\cite{KKMC}. 
{For the signal MC sample,} we generate 
$\tau^+\tau^-$ {pairs}, where 
{one} $\tau$ decays into  
{a lepton and two charged mesons,}
{using {a three-body {phase space} model,}}
and the other $\tau$ decays generically.
Other {backgrounds,} including
$B\bar{B}$ and {continuum} 
$e^+e^-\to q\bar{q}$ ($q=u,d,s,c$) events, Bhabha events, 
and two-photon processes are generated by 
EvtGen~\cite{evtgen},
BHLUMI~\cite{BHLUMI}, 
and
{AAFH~\cite{AAFH}}, respectively. 
{The event selection is optimized {mode-by-mode}
since the {backgrounds} are mode dependent.}
All kinematic variables are calculated in the laboratory frame
unless otherwise specified.
In particular,
variables
calculated in the $e^+e^-$ center-of-mass (CM) system
are indicated by the superscript ``CM''.

\section{Event Selection}

%
%
{Since the majority of $\tau$ decays 
{produce} {one-prong final} states~\cite{PDG},}
{we search for $\tau^+\tau^-$ events in which one $\tau$ 
(the {signal $\tau$}) decays into {a} lepton
and
two charged mesons ($\pi^\pm$ or $K^\pm$)}
and the other $\tau$~(the {tag $\tau$}) 
decays 
into  one charged track 
{with} any number of additional 
photons and neutrinos. 
Candidate $\tau$-pair events are required to have 
four tracks with {zero} net charge.

%
%

{We start by reconstructing }
four charged tracks and any number of photons within the fiducial volume
 defined by $-0.866 < \cos\theta < 0.956$,
{where $\theta$ is 
the polar angle {relative} to 
the direction opposite to 
that of 
the {incident} $e^+$ beam in 
{the} laboratory frame.}
The transverse momentum ($p_t$) of each charged track
and {the} energy of each photon ($E_{\gamma}$) 
are 
{required to satisfy} $p_t> $ 0.1 GeV/$c$ and $E_{\gamma}>0.1$ GeV,
respectively.
{For each charged track, 
the distance of the closest point with 
respect to the interaction point 
is required to be 
less than 0.5 cm in the transverse direction 
and less than 
3.0 cm in the longitudinal direction.}

%
%

{Using the plane perpendicular to the CM
thrust axis~\cite{thrust},
which is calculated from 
the observed tracks and photon 
{candidates,}
we separate the particles in an event
into two hemispheres.
These  are referred to as the signal and 
tag sides. 
The tag side contains one charged track}
while the signal side contains three charged tracks.
We require one charged track 
{on}  the signal side 
{to be} 
identified as a lepton.
The lepton identification criteria are 
${\cal P}(\ell) > 0.95$ 
and
{the momentum thresholds} 
{are}
{listed} in Table~\ref{tbl:thrust}.
The electron (muon) identification
{efficiency
is} 91\% (85\%)
while
{the probability to misidentify {a} pion
as
{an} {electron} ({a} muon)}
is below 0.5\% (2\%).
In order to take into account the emission
of  bremsstrahlung photons from the electron,
the momentum of {each} 
electron track
is reconstructed by 
{adding
the momentum of every photon
within} 
0.05 rad along
{the track.}
To reduce generic $\tau^+\tau^-$ 
and $q\bar{q}$ background events,
{we veto events that have a photon 
{on} the signal side.}

%
%

To ensure that the missing particles are neutrinos rather
than photons or charged particles that pass  outside the detector acceptance,
we impose requirements on the missing 
momentum $\vec{p}_{\rm miss}$,
which 
is
calculated by subtracting the
vector sum of the momenta of all tracks and photons
from the sum of the $e^+$ and $e^-$ beam momenta.
We require that the magnitude of $\vec{p}_{\rm miss}$
be  greater than 1.0 GeV/$c$,
and {that} its direction point into the fiducial volume of the
detector.
Furthermore,
we reject the event if the  
direction of the missing momentum 
{traverses the
gap between the barrel and endcap of the ECL}. 
Since neutrinos are 
emitted only 
{on} the tag side,
the direction of
$\vec{p}_{\rm miss}$
should lie within the tag side of the event.
The cosine of the
opening angle between
$\vec{p}_{\rm miss}$
and the charged track 
{on} the tag side 
{in the CM system,}
$\cos \theta^{\mbox{\rm \tiny CM}}_{\rm tag-miss}$, 
{{should be} in the range 
$0.4<\cos \theta^{\mbox{\rm \tiny CM}}_{\rm tag-miss}<0.98$.

%
%

{The remaining two tracks on
the signal side are identified 
as $K^\pm$ ($\pi^\pm$) if they satisfy
the condition  ${\cal{P}}(K/\pi) > 0.8$ ($<0.4$).
If either track has 
{a value in the intermediate range},
$0.4 < {\cal{P}}(K/\pi) < 0.8$,
the event is rejected.}
The kaon (pion) identification
{efficiency
is} 80\% (88\%)
while
{the probability to misidentify {a} pion
(kaon)
as
{a} {kaon} ({a} pion)}
is below 10\% (12\%).
{In order to 
{reduce background from mesons reconstructed}
from {photon} conversions
(i.e. $\gamma \rightarrow e^+e^-$),}
we require that two charged meson candidates 
have ${\cal{P}}(e) <0.1$.
Furthermore, we require 
${\cal{P}(\mu)} <0.1$
to suppress 
{the} two-photon background 
{process} $e^+e^- \to e^+e^-\mu^+\mu^-$.

%

To reject $q\bar{q}$ background,
{we require the magnitude of the thrust ($T$)
to be in the ranges} 
{given} in Table \ref{tbl:thrust} 
(see Fig.~\ref{cut1} (a) and Fig.~\ref{cut2} (a)). 
We also require $5.5$ GeV 
$< E^{\mbox{\rm{\tiny{CM}}}}_{\rm{vis}} < 10.0$ GeV, 
where $E^{\mbox{\rm{\tiny{CM}}}}_{\rm{vis}}$ 
is the total visible energy in the CM system, defined as 
the sum of the energies of {the} lepton, two charged mesons,
the charged track 
{on} the tag side (with a pion mass hypothesis)
and all photon {candidates} 
(see Fig.~\ref{cut1} (b)).
The 
{invariant mass reconstructed 
{from the charged track and any photons}}
{on} the tag side 
$m_{\rm tag}$, 
is required to be less than 1.00 GeV/$c^2$
(see Fig.\ref{cut2} (b)).
In order to reduce $q\bar{q}$ background,
{a kaon veto}
{${\cal P}(K/\pi) < 0.8$}
is applied to the lepton
and 
tracks {on} the tag side 
for the $\mu\pi K$ and $\mu KK$ modes.

We remove events 
if 
{$K^0_S$} candidates are 
reconstructed
from two {oppositely-charged tracks 
{on} the signal side
{with}
{an invariant mass}
{0.470 GeV/$c^2 < M_{\pi^+\pi^-} <0.525$ {GeV/$c^2$},
{assuming the pion} {mass} for both tracks,}}
and the $\pi^+\pi^-$ vertex is
displaced from the interaction point (IP)
in the direction of the pion pair momentum~\cite{cite:ks}.
{Events} including a 
{$K^0_L$} meson also 
{constitute}
background since 
{the undetected} 
{$K^0_L$} results 
{in fake missing momentum.}
Therefore, we veto events 
{with}
{$K^0_L$} candidates, 
which are selected from 
hit clusters in {the} KLM that are not associated 
with either an ECL cluster
or with a charged track~\cite{cite:kl},
for the $\mu hh'$ modes.

To suppress the $B\bar{B}$ and $q\bar{q}$ background,
{we require} that
the number of photons 
{on} the tag side $n_{\gamma}^{\rm{TAG}}$ 
be $n_{\gamma}^{\rm{TAG}} \leq 2$ 
and $n_{\gamma}^{\rm{TAG}}\leq 1$
for hadronic and leptonic tag 
{decays,} respectively (see Fig.~\ref{cut3}).
For all kinematic distributions shown 
in {Figs.~\ref{cut1}, \ref{cut2} and \ref{cut3}},
reasonable agreement between the data and background MC is observed.

To reduce two-photon background, 
we apply 
{an} electron veto ${\cal P}(e) < 0.1$  to 
{the track}  
{on} the tag side 
for the $e\pi\pi$ and $e\pi K$ modes.
Furthermore, 
we require that the momentum of the electron and 
track 
{on} the tag side
in the CM system be less than 4.5 {GeV/$c$} 
to reduce 
{Bhabha background} in the $e\pi\pi$ modes.

\begin{table}
\caption{Selection criteria for 
lepton momentum ($p_\ell$) 
and magnitude of thrust ($T$).}
\label{tbl:thrust}
\begin{tabular}{|c|c|c|}\hline \hline 
Mode & $p_\ell$  {(GeV/$c$)} & $T$  \\\hline
$\tau\to\mu\pi\pi$ &  $p_\mu > 1.4$ &
$0.90 < T < 0.97$ \\
$\tau\to\mu K\pi$ &  $p_\mu > 1.1$  &
$0.92 < T < 0.98$ \\
$\tau\to\mu KK$ &  $p_\mu > 0.8$  &
$0.92 < T < 0.98$ \\
$\tau\to e\pi\pi$ &  $p_e > 0.6$   &
$0.90 < T < 0.97$ \\
$\tau\to eK\pi$  &$p_e > 0.4$  &
$0.90 < T < 0.97$ \\
$\tau\to eKK$  &$p_e > 0.4$  &
$0.90 < T < 0.98$ \\\hline\hline
\end{tabular}
\end{table}

%
%

\begin{figure}
\begin{center}
       \resizebox{0.8\textwidth}{0.4\textwidth}{\includegraphics
        {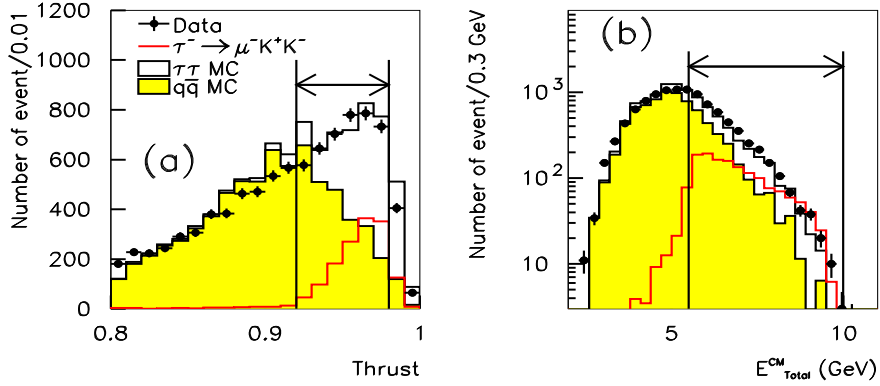}}
 \caption{
 Distribution of 
(a) the magnitude of thrust
and
(b) the total visible energy in the CM system.
While the signal MC  ($\tau^-\to\mu^-K^+K^-$)
 distribution is normalized arbitrarily, 
 the data and background MC are normalized to the same luminosity.
{The selected} regions are indicated
 by the
 arrows.}
\label{cut1}
\end{center}
\end{figure}

\begin{figure}
\begin{center}
       \resizebox{0.8\textwidth}{0.4\textwidth}{\includegraphics
        {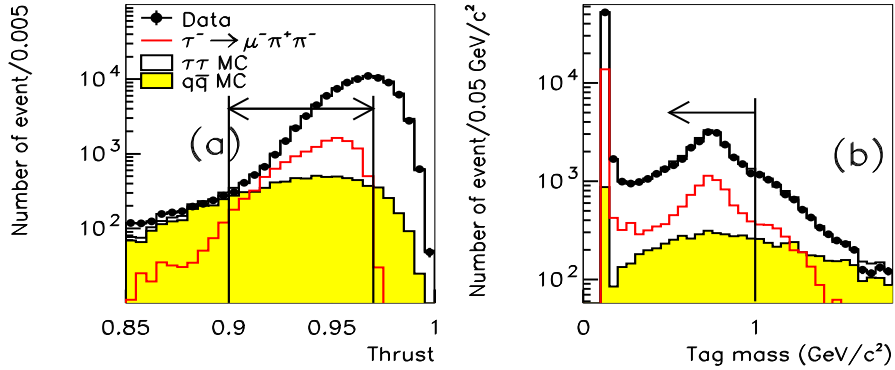}}
 \caption{
 Distribution of 
(a) the magnitude of thrust and 
(b) invariant mass using
particles {on} the tag side.
While the signal MC  ($\tau^-\to\mu^-\pi^+\pi^-$)
 distribution is normalized arbitrarily, 
 the data and background MC are normalized to the same luminosity.
 {The selected} 
 regions are indicated
 by the
 arrows.}
\label{cut2}
\end{center}
\end{figure}

\begin{figure}
\begin{center}
       \resizebox{0.8\textwidth}{0.4\textwidth}{\includegraphics
 {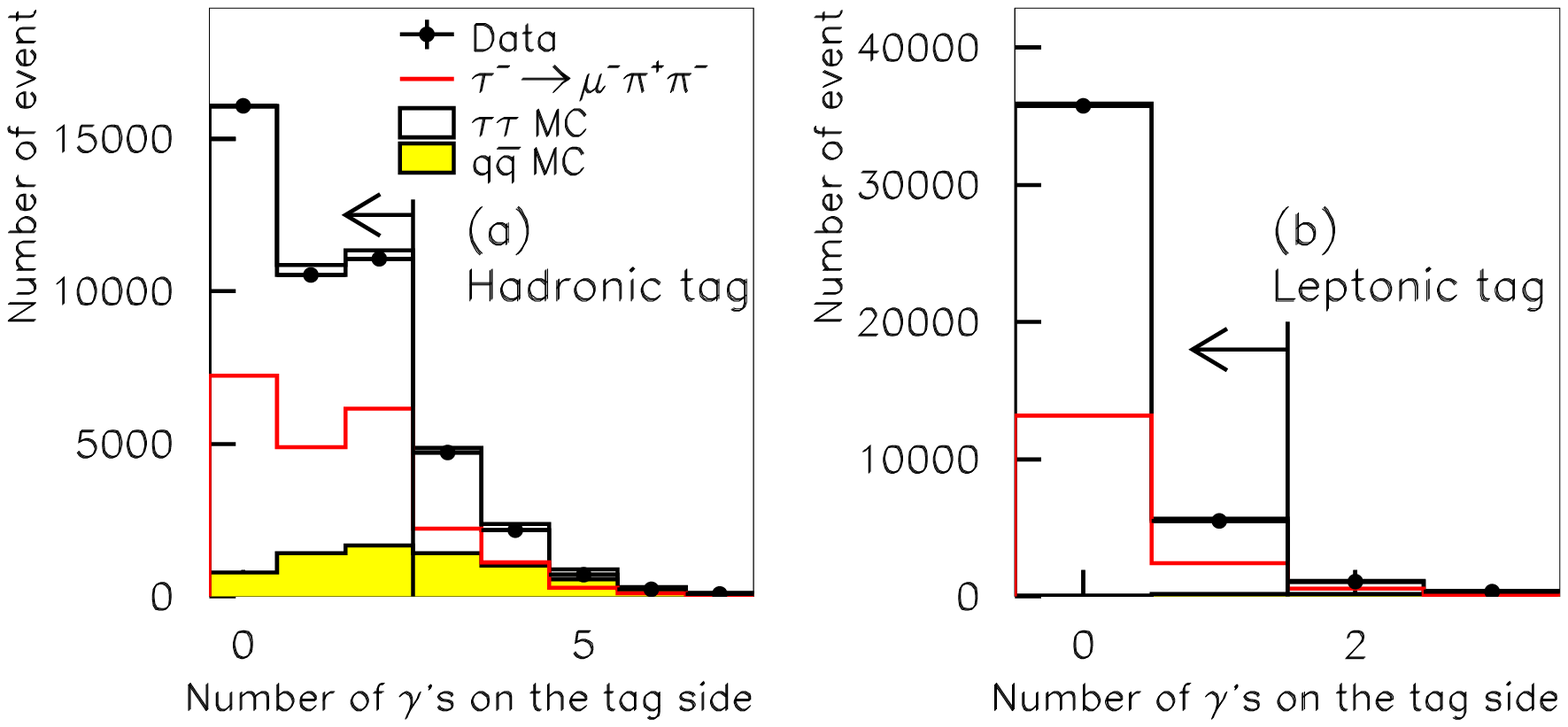}}
 \caption{
 Distributions of the number of 
{photons} {on} the tag side 
 for (left) hadronic and (right) leptonic {tags.}
 While the signal MC ($\tau^-\to\mu^-\pi^+\pi^-$)
 distribution is normalized arbitrarily, 
 the data and background MC are normalized to the same luminosity.
{The selected} regions are indicated
 by the
 arrows.}
\label{cut3}
\end{center}
\end{figure}

%
%
Finally, 
to suppress  backgrounds from generic 
$\tau^+\tau^-$ and $q\bar{q}$ events, 
we apply a selection based on the magnitude of the missing momentum 
${p}_{\rm{miss}}$ and {the}
missing mass squared $m^2_{\rm{miss}}$.
{We apply different selection criteria depending on the lepton
identification of the charged track 
{on} the tag side; 
{two neutrinos are emitted}
if the track is an electron or muon 
(leptonic tag) while  {only one is emitted} 
if the track is a hadron (hadronic tag).}
For the $ehh'$, $\mu\pi\pi$ and $\mu KK$ modes,
we require
the following relation between
$p_{\rm{miss}}$ and $m^2_{\rm{miss}}$:
$p_{\rm{miss}} > -7.0\times m^2_{\rm{miss}}-1$
and 
$p_{\rm{miss}} > 7.0\times m^2_{\rm{miss}}-1.0$
for {the} hadronic tag
and 
$p_{\rm{miss}} > -8.0\times m^2_{\rm{miss}}+0.2$
and 
$p_{\rm{miss}} > 1.8\times m^2_{\rm{miss}}-0.4$
for {the} leptonic tag,
where $p_{\rm{miss}}$ is in GeV/$c$ and
$m_{\rm{miss}}$ is in GeV/$c^2$
(see Fig. \ref{fig:pmiss_vs_mmiss2}).
For the $\mu\pi K$ modes,
we require
the following relation between
$p_{\rm{miss}}$ and $m^2_{\rm{miss}}$:
$p_{\rm{miss}} > -8.0\times m^2_{\rm{miss}}-0.5$
and 
$p_{\rm{miss}} > 8.0\times m^2_{\rm{miss}}-0.5$
for {the} hadronic tag
and 
$p_{\rm{miss}} > -9.0\times m^2_{\rm{miss}}+0.4$
and 
$p_{\rm{miss}} > 1.8\times m^2_{\rm{miss}}-0.4$
for {the} leptonic tag.

\begin{figure}
\begin{center}
 \resizebox{0.7\textwidth}{0.7\textwidth}{\includegraphics
 {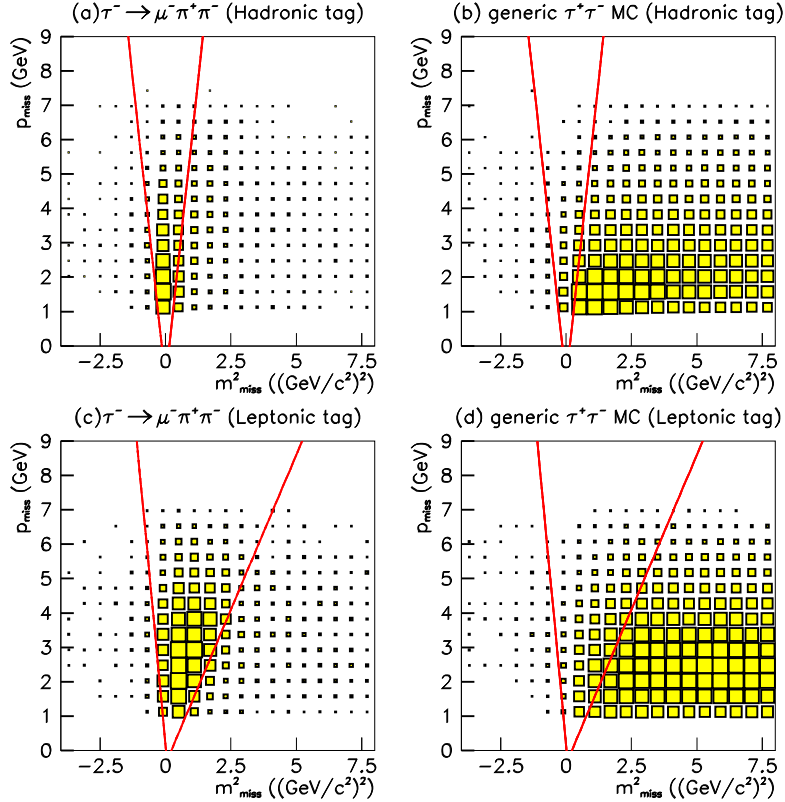}}
 \vspace*{-0.5cm}
 \caption{
Scatter-plots of
{$p_{\rm miss}$ 
{vs.} 
$m_{\rm miss}^2$:
(a) and (b) 
show
the signal MC 
($\tau^-\to\mu^-\pi^+\pi^-$)
and
generic $\tau^+\tau^-$ MC 
distributions,
respectively,
for the hadronic tags
while (c) and (d)  
{are}
the same distributions 
for the {leptonic tags.}}
{The selected} regions are indicated by lines.}
\label{fig:pmiss_vs_mmiss2}
\end{center}
\end{figure}

\section{Signal and Background Estimation}

The signal candidates are examined in the 
two-dimensional {plot} of the $\ell hh'$ invariant
mass~($M_{{\ell hh'}}$) {versus} 
the difference of their energy from the 
beam energy in the CM system~($\Delta E$).
A signal event should have $M_{{\ell hh'}}$
close to the $\tau$-lepton mass ($m_{\tau}$) and
$\Delta E$ close to zero.
For all modes,
the $M_{{\ell hh'}}$ and $\Delta E$  resolutions are parameterized
from fits to the signal MC {distributions,}  
with  an asymmetric Gaussian function that takes into account 
{initial-state} radiation.
The resolutions 
{in} $M_{\ell hh'}$ and $\Delta E$ 
are 
{listed} in Table~\ref{tbl:width}.

\begin{table}
\begin{center}
\caption{Summary {of} $M_{\rm \ell hh'}$  
and
$\Delta E$
{resolutions} ($\sigma^{\rm{high/low}}_{M_{\rm{\ell hh'}}}$ (MeV/$c^2$) 
and 
$\sigma^{\rm{high/low}}_{\Delta E}$ (MeV)).
Here $\sigma^{\rm high}$ ($\sigma^{\rm low}$)
means the standard deviation
{on the} higher (lower) side of the peak.}
\label{tbl:width}
\vspace*{0.2cm}
\begin{tabular}{c|cccc} \hline\hline
Mode
& $\sigma^{\rm{high}}_{M_{\rm{\ell hh'}}}$
& $\sigma^{\rm{low}}_{M_{\rm{\ell hh'}}}$ 
& $\sigma^{\rm{high}}_{\Delta E}$ 
&  $\sigma^{\rm{low}}_{\Delta E}$ 
 \\ \hline
$\tau^-\to\mu^-\pi^+\pi^-$
& 4.8  &  5.5 & 13.7 & 18.0 \\
$\tau^-\to\mu^+\pi^-\pi^-$
& 5.3   &  5.4 & 14.1 & 18.8 \\
$\tau^-\to e^-\pi^+\pi^-$
& 5.3  &  5.9 & 14.7 & 21.2 \\
$\tau^-\to e^+\pi^-\pi^-$
& 5.6  &  5.9 & 14.2 & 21.3 \\
$\tau^-\to\mu^- K^+K^-$
& 3.6  & 4.0 & 11.4 & 18.0 \\
$\tau^-\to\mu^+ K^-K^-$
& 3.4  &  3.6 & 11.4 & 18.8 \\
$\tau^-\to e^- K^+K^-$
& 4.0  &  4.3 & 13.7 & 20.5 \\
$\tau^-\to e^+ K^-K^-$
& 3.5  &  4.5 & 13.9 & 21.3 \\
$\tau^-\to\mu^-\pi^+K^-$
& 4.5  &  5.0 & 13.6 & 18.6 \\
$\tau^-\to  e^-\pi^+K^-$
& 4.7  &  5.4 & 13.6 & 21.7 \\
$\tau^-\to\mu^-K^+\pi^-$
& 4.6  &  5.1 & 14.3 & 18.3 \\
$\tau^-\to  e^-K^+\pi^-$
& 4.6  &  5.5 & 13.5 & 21.6 \\
$\tau^-\to\mu^+K^-\pi^-$
& 4.5  & 5.0 & 12.4 & 18.6 \\
$\tau^-\to  e^+K^-\pi^-$
& 4.9  &  5.3 & 13.0 & 20.8 \\
\hline\hline
\end{tabular}
\end{center}
\end{table}

{{To evaluate the branching {fractions,}
we use  {elliptical signal regions}
{that} {contain} 90\%
of the MC signal {events satisfying} all {selection criteria.}
We blind the data in the signal region
{until all selection criteria are finalized}
so as not to bias our choice of selection criteria. }}

{For the $ehh'$ modes the dominant background 
{is}} 
from
two-photon 
{processes; the} 
fraction of $q\bar{q}$ and generic $\tau^+\tau^-$ events 
is small due to 
{the}  low electron fake rate.
{For the $\mu\pi\pi$ mode
the dominant background 
{is}
from}
$q\bar{q}$ {processes}
{and a smaller} background 
{is}
from
generic
$\tau^+\tau^-$ events in the $\Delta E<$ 0 GeV and 
$M_{\mu\pi\pi} < $ $m_{\tau}$ region,
which 
are combinations of a fake muon and two pions.
{For the $\mu\pi K$ mode,
the dominant background 
{is}
from
generic
$\tau^+\tau^-$ events
{that}
are
combinations of 
{a}
fake muon, {a} fake kaon and 
{a}
{true} pion.}
If a pion is misidentified as a kaon,
the reconstructed mass from
generic $\tau^+\tau^-$ background 
{can}
be greater than the $\tau$ lepton mass 
{because of the larger kaon mass.}
For the $\mu KK$ mode,
{the dominant background 
{originates}} from
$q\bar{q}$ events  
and
$\tau^+\tau^-$ events.

{The number of 
{background events
in the signal region}
is {estimated from} 
the data
{remaining} after event selection
in {the} sideband region.
For {the} $e hh'$ and  $\mu  KK$ modes, 
since the number of remaining data events is small,
the number of background events in the signal region
is estimated 
by interpolating the number of observed events in the sideband {region}
{defined as} 
the {range}
{$\pm 20\sigma_{M_{\ell hh'}}$} and $\pm 5\sigma_{\Delta E}$
excluding the signal region,
assuming that the background distribution is uniform in the sideband
region.
For the $\mu\pi\pi$ and $\mu\pi K$ modes, 
we {estimate}
the number of background events in the signal region
by fitting to observed data in {the}
sideband region
using 
{a} probability density function (PDF)
{that describes} the shapes of 
{the} background distributions
along the $M_{\mu\pi\pi}$ axis within $\pm 5\sigma_{\Delta E}$.
For the $\mu\pi\pi$ {mode},
{the} PDFs of generic $\tau\tau$ and $q\bar{q}$ events
are determined using MC samples,
assuming exponential 
and first-order polynomial {distributions,} 
{respectively (see Fig.~\ref{fig:mupipi})}.}
For the $\mu\pi K$ modes, 
we {parameterize}
{the PDF}
by {a} 3rd-order polynomial function
{that is fitted to}
the 
data {remaining} 
in 
{the} sideband region.
The signal efficiency and 
the number of expected background 
events in {the} signal region for each mode 
are summarized in Table~\ref{tbl:eff}.

The dominant systematic uncertainties
for this analysis
{come
from tracking efficiencies
and particle identification.}
The uncertainty due to the charged track finding is 
estimated to be 1.0\% per charged 
{track; the} 
total uncertainty due to the charged track finding is
4.0\%.
{The uncertainties due to lepton {identification} 
{are} 
2.2\% and 1.9\% for electron and muon, respectively.}
{The uncertainty due to pion and kaon {identification} 
{is} 1.3\% and 1.8\% per pion and kaon, respectively. }
The uncertainty due to the $e$-veto 
{on} the tag side applied for
{the}
$\tau\to e\pi\pi$  and $\tau\to e\pi K$  
modes is estimated as the uncertainty 
{in} the electron identification
times the branching 
{fraction} of 
{$\tau^-\to e^-\bar{\nu}_e\nu_{\tau}$} (0.4\%).
The uncertainties due to MC statistics and luminosity
are estimated to be 
{(2.5 $-$ 3.4)\%} and 1.4\%, respectively.
The uncertainty due to the trigger efficiency is negligible 
compared 
{to} the other uncertainties.
All these uncertainties are added in {quadrature}
{giving}
total systematic uncertainties for  all modes 
{in the (5.9 $-$ 6.8)\% range.}

\begin{table}
\begin{center}
\caption{ The signal efficiency~($\varepsilon$), 
the number of expected background {events}~($N_{\rm BG}$)
estimated from the  sideband data, 
{the} {total} 
systematic uncertainty~($\sigma_{\rm syst}$),
{the} number of observed events 
in the signal region~($N_{\rm obs}$), 
90\% C.L. upper limit on the number of signal events including 
systematic uncertainties~($s_{90}$) 
and 90\% C.L. upper limit on 
the branching  fraction 
for each individual mode. }
\label{tbl:eff}
\begin{tabular}{c|cccccc}\hline \hline
Mode &  $\varepsilon$~{(\%)} & 
$N_{\rm BG}$  & $\sigma_{\rm syst}$~{(\%)}
& $N_{\rm obs}$ & $s_{90}$ & 
${\cal{B}}~(10^{-8})$ \\ \hline
$\tau^-\to \mu^-\pi^+\pi^- $ & 3.69 & $1.12\pm{0.38}$ & 5.9
 & 0 & 1.53  & 3.3\\
$\tau^-\to \mu^+\pi^-\pi^- $ & 3.84 & $0.73\pm{0.25}$ & 5.9
 & 0 & 1.77  & 3.7\\
$\tau^-\to e^-\pi^+\pi^- $ & 3.99 & $0.34\pm{0.15}$ & 6.0
 & 0 & 2.15  & 4.4\\
$\tau^-\to e^+\pi^-\pi^- $ & 3.91 & $0.10\pm{0.07}$ & 6.0
 & 1 & 4.21  & 8.8\\

$\tau^-\to \mu^-K^+K^- $ & 2.40 & $0.52\pm{0.23}$ & 6.7
 & 0 & 1.92  & 6.8\\
$\tau^-\to \mu^+K^-K^- $ & 2.07 & $0.00\pm{0.06}$ & 6.8
 & 0 & 2.46  & 9.6\\
$\tau^-\to e^-K^+K^- $ & 3.50 & $0.11\pm{0.08}$ & 6.5
 & 0 & 2.35  & 5.4\\
$\tau^-\to e^+K^-K^- $ & 3.28 & $0.05\pm{0.05}$ & 6.6
 & 0 & 2.43  & 6.0\\

$\tau^-\to \mu^-\pi^+K^- $ & 2.63 & $0.67\pm{0.14}$ & 6.3
 & 2 & 5.05  & 16\\
$\tau^-\to  e^-\pi^+K^- $ & 3.02 & $0.33\pm{0.19}$ & 6.4
 & 0  & 2.12  & 5.8 \\

$\tau^-\to \mu^-K^+\pi^- $ & 2.60 & $1.04\pm{0.32}$ & 6.3
 & 1 & 3.34  & 10\\
$\tau^-\to  e^-K^+\pi^- $ & 2.98 & $0.57\pm{0.19}$ & 6.4
 & 0 & 1.90  & 5.2\\

$\tau^-\to \mu^+K^-\pi^- $ & 2.61 & $1.37\pm{0.21}$ & 6.3
 & 1 & 3.16  & 9.4 \\
$\tau^-\to  e^+K^-\pi^- $ & 2.83 & $0.10\pm{0.07}$ & 6.4
 & 0 & 2.40  & 6.7\\

\hline\hline
\end{tabular}
\end{center}
\end{table}

\section{Upper Limits on the branching fractions}

Finally, 
{we examine the data in 
the signal region}
and observe two candidate events for the $\mu^- \pi^+K^-$ mode,
{one} candidate event for each of the $\mu^-K^+\pi^-$,
$\mu^+ \pi^-K^-$ and 
$e^+ \pi^-\pi^-$  modes,
and no candidate events for 
{the} other modes.
{These numbers of events are consistent} with
{the} expected numbers of background events.
Since no statistically significant excess of data over
the expected background {is} observed,
we set the following upper limits on the branching fractions 
of $\tau\to\ell hh'$
based on the Feldman-Cousins method~\cite{cite:FC}.
The 90\% C.L. upper limit on the number of signal events 
including  a systematic uncertainty~($s_{90}$) is obtained 
{using} the POLE program without conditioning~\cite{pole}
{based on}
the number of expected {background events}, 
{the number of observed events}
and the systematic uncertainty.
The upper limit on the branching fraction ($\cal{B}$) is then given by
\begin{equation}
{{\cal{B}}(\tau\to\ell hh') <
\displaystyle{\frac{s_{90}}{2N_{\tau\tau}\varepsilon{}}}},
\end{equation}
where $N_{\tau\tau}$ is the number of $\tau^+\tau^-$ pairs, and 
$\varepsilon$ is the signal efficiency.
{The value {$N_{\tau\tau} =  616.6\times 10^6$}} is obtained 
from 
{the} integrated luminosity times 
the cross section 
{for} {$\tau$-pair production,} which 
is calculated 
{in the updated version of 
KKMC~\cite{tautaucs} to be 
$\sigma_{\tau\tau} = 0.919 \pm 0.003$ nb.}
Table~\ref{tbl:eff}
{summarizes} 
information about 
{the}
upper limits 
for all modes.
{We} obtain the following 90\% C.L. upper limits 
{on the branching fractions:} 
${\cal{B}}(\tau\rightarrow ehh')
 < (4.4-8.8)\times 10^{-8}$
and 
${\cal{B}}(\tau\rightarrow \mu hh')< (3.3-16)\times 10^{-8}$.
These results improve {upon} 
previously published upper limits
by factors {of} 1.6 to 8.8~\cite{lhh_belle}.

\begin{figure}
\begin{center}
       \resizebox{0.4\textwidth}{0.4\textwidth}{\includegraphics
 {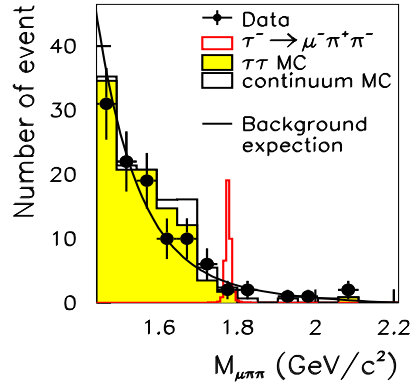}}
 \caption{
Mass distribution of $\mu^-\pi^+\pi^-$ 
within the $\pm 5\sigma_{\Delta E}$ region after event selection.
While the signal MC  ($\tau^-\to\mu^-\pi^+\pi^-$)
 distribution is normalized arbitrarily, 
 the data and background MC are normalized to the same luminosity.
The expected background is shown as the solid {histogram.} }
\label{fig:mupipi}
\end{center}
\end{figure}

\section{Summary}
We have searched for {lepton-flavor and lepton-number-violating} $\tau$ decays 
into 
{a} 
lepton and two charged mesons ($h,h' = \pi^\pm$ or $K^\pm$)
using 671 fb$^{-1}$ of data.
{We found no excess of signal in 
{any of the modes.}
The 
{resulting} 90\% C.L. upper limits 
{on the branching fractions,} 
{${\cal{B}}(\tau\rightarrow ehh')
 < (4.4-8.8)\times 10^{-8}$}
and 
${\cal{B}}(\tau\rightarrow \mu hh')< (3.3-16)\times 10^{-8}$,
improve {upon} 
previously published results
by factors {of} 1.6 to 8.8.}
These more stringent upper limits can be used
to constrain the  
{parameter spaces} in various models of new physics.

\begin{figure}
\begin{center}
       \resizebox{0.35\textwidth}{0.35\textwidth}{\includegraphics
 {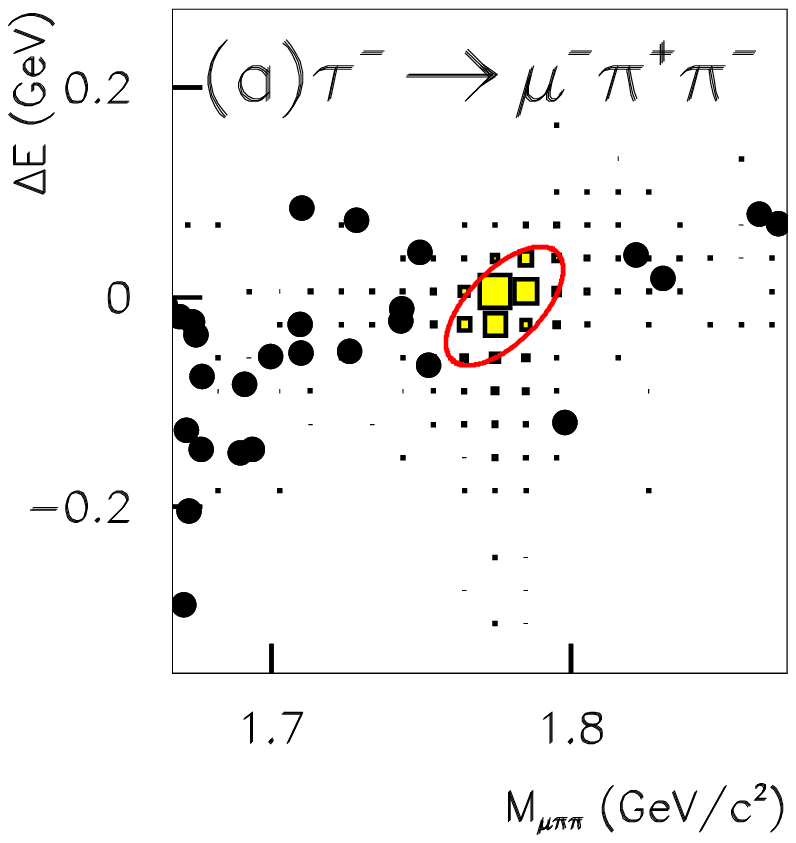}}
\hspace*{-0.8cm}
       \resizebox{0.35\textwidth}{0.35\textwidth}{\includegraphics
 {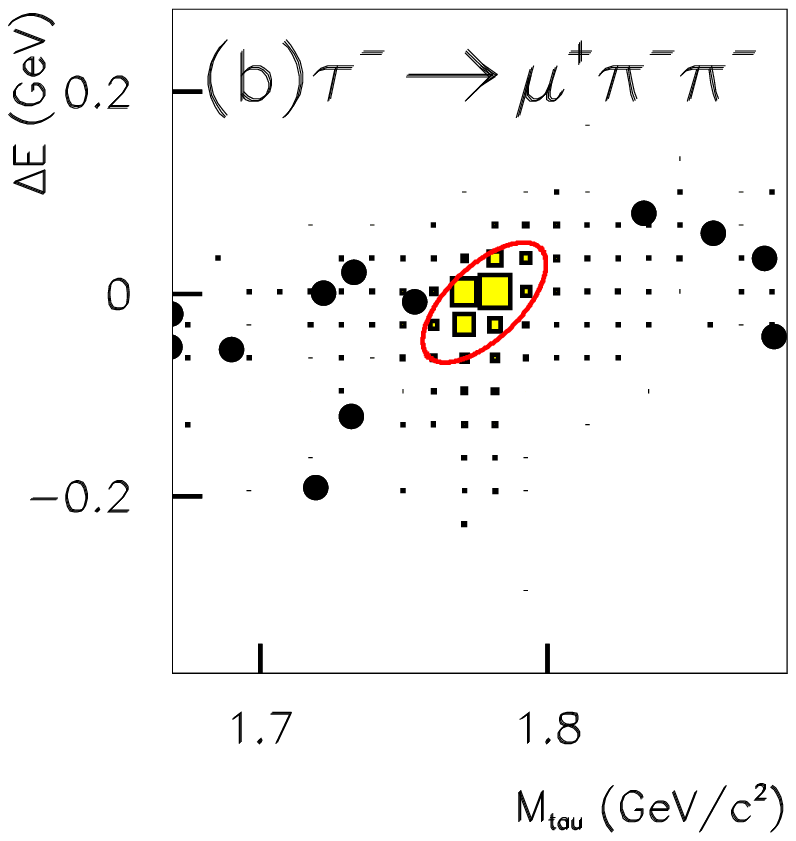}}
\hspace*{-0.8cm}
       \resizebox{0.35\textwidth}{0.35\textwidth}{\includegraphics
 {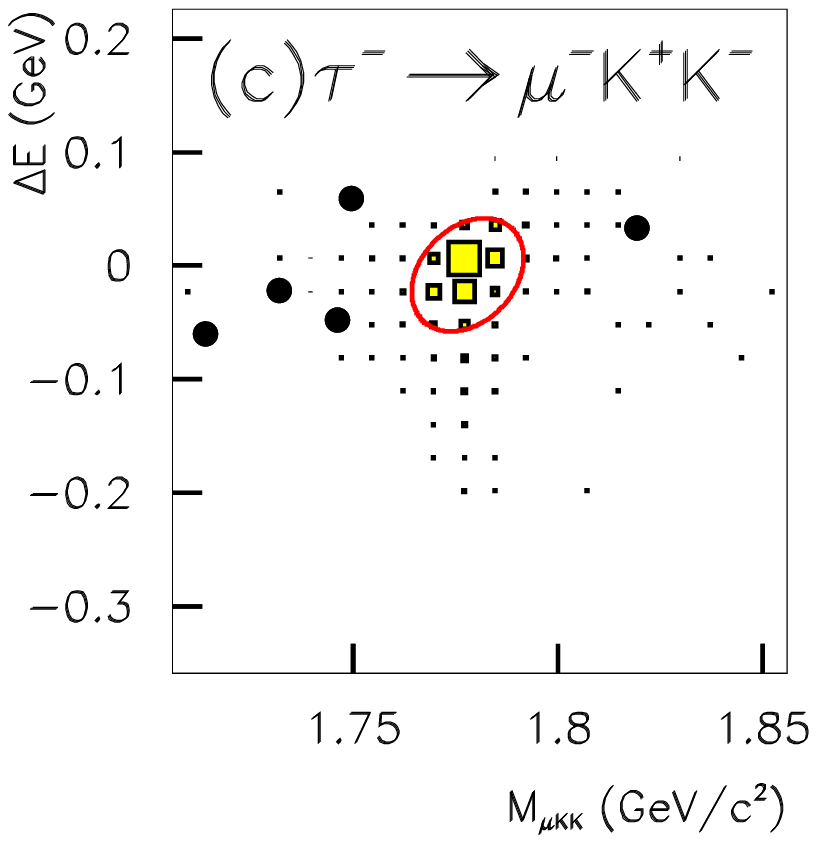}}
       \resizebox{0.35\textwidth}{0.35\textwidth}{\includegraphics
 {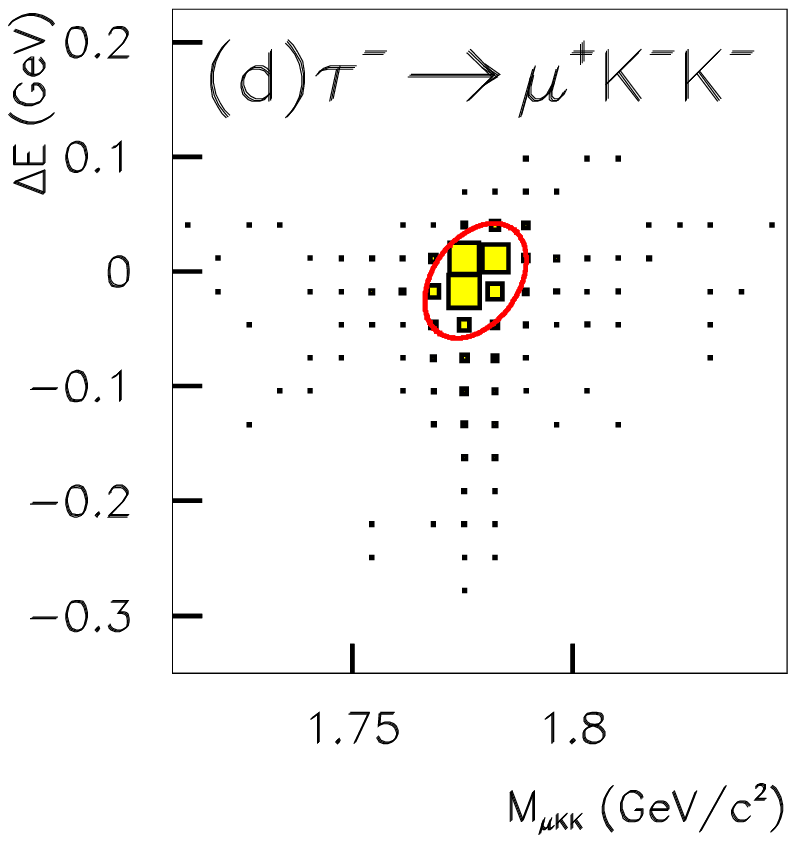}}
\hspace*{-0.8cm}
       \resizebox{0.35\textwidth}{0.35\textwidth}{\includegraphics
 {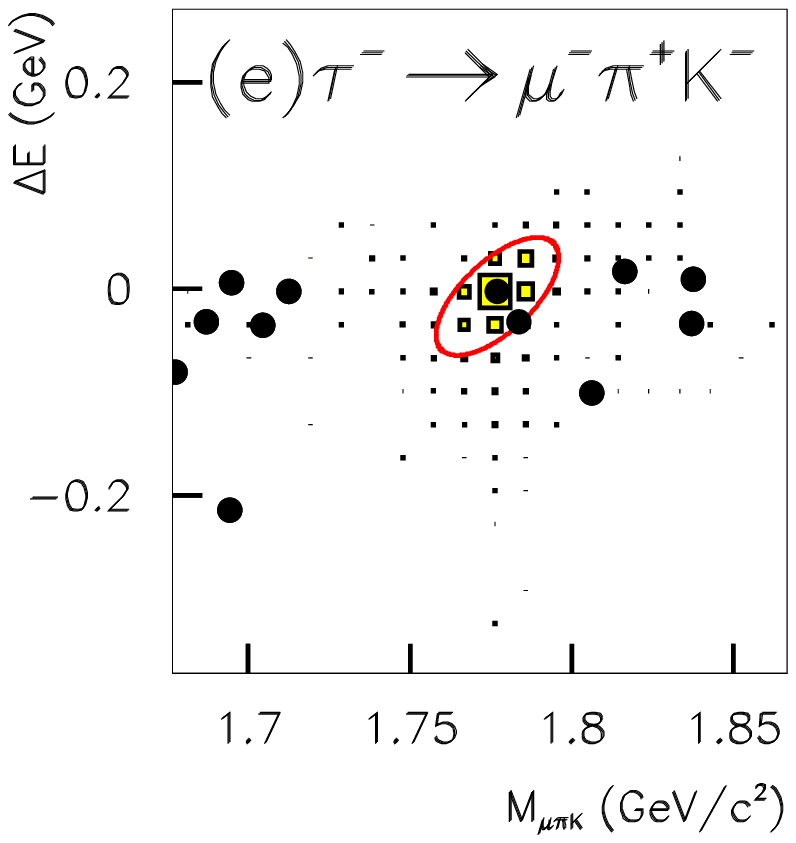}}
\hspace*{-0.8cm}
       \resizebox{0.35\textwidth}{0.35\textwidth}{\includegraphics
 {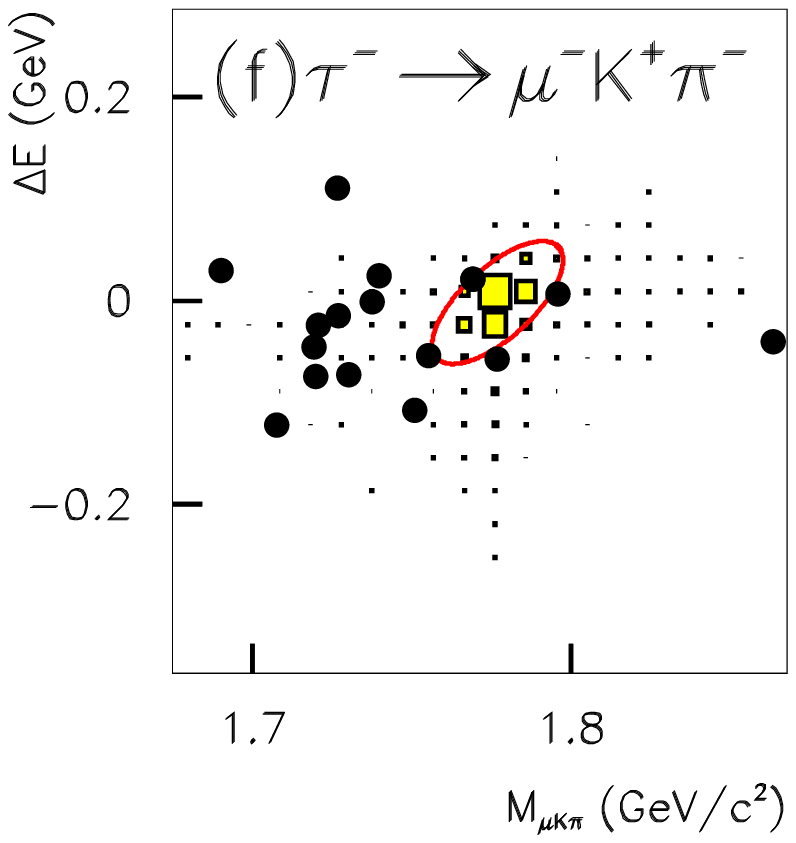}}
       \resizebox{0.35\textwidth}{0.35\textwidth}{\includegraphics
 {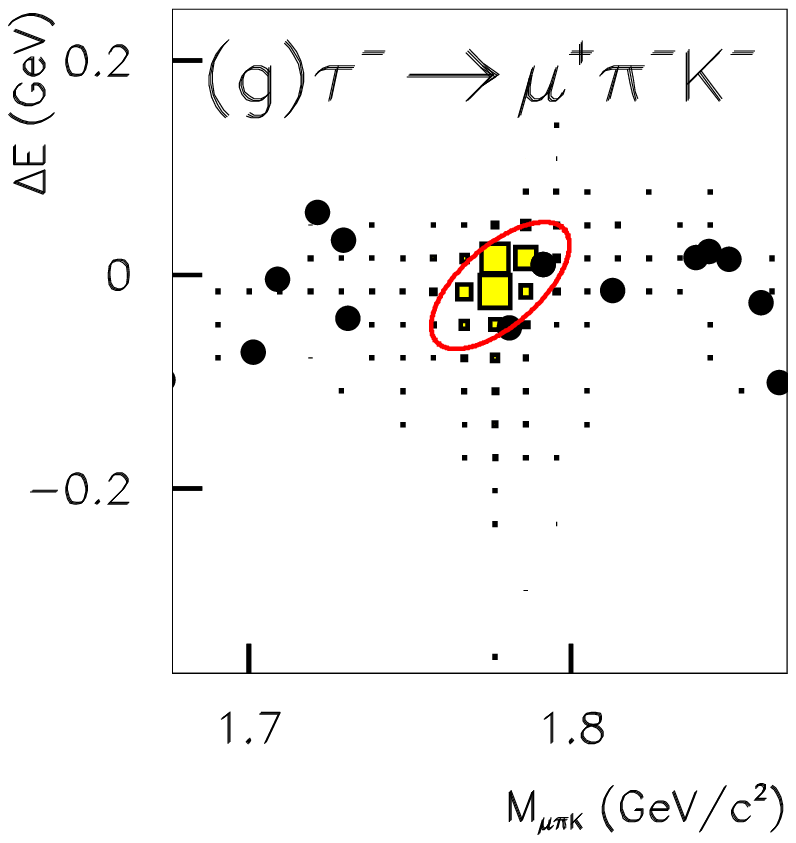}}
 \caption{
Scatter-plots in the $M_{\ell hh'}$ -- $\Delta{E}$ plane  
{within}
the $\pm 20 \sigma$ area for the
(a) $\tau^-\rightarrow \mu^-\pi^+\pi^-$,
(b) $\tau^-\rightarrow \mu^+\pi^-\pi^-$,
(c) $\tau^-\rightarrow \mu^-K^+K^-$,
(d) $\tau^-\rightarrow \mu^+K^-K^-$,
(e) $\tau^-\rightarrow \mu^-\pi^+K^-$,
(f) $\tau^-\rightarrow \mu^-K^+\pi^-$,
and
(g) $\tau^-\rightarrow \mu^+\pi^-K^-$
modes.
The data are indicated by the solid circles.
The filled boxes show the MC signal distribution
with arbitrary normalization.
The elliptical signal
{regions}
shown by {the} solid 
{curves}
are used for evaluating the signal yield.
}
\label{fig:ehh}
\end{center}
\end{figure}

\begin{figure}
\begin{center}
       \resizebox{0.35\textwidth}{0.35\textwidth}{\includegraphics
 {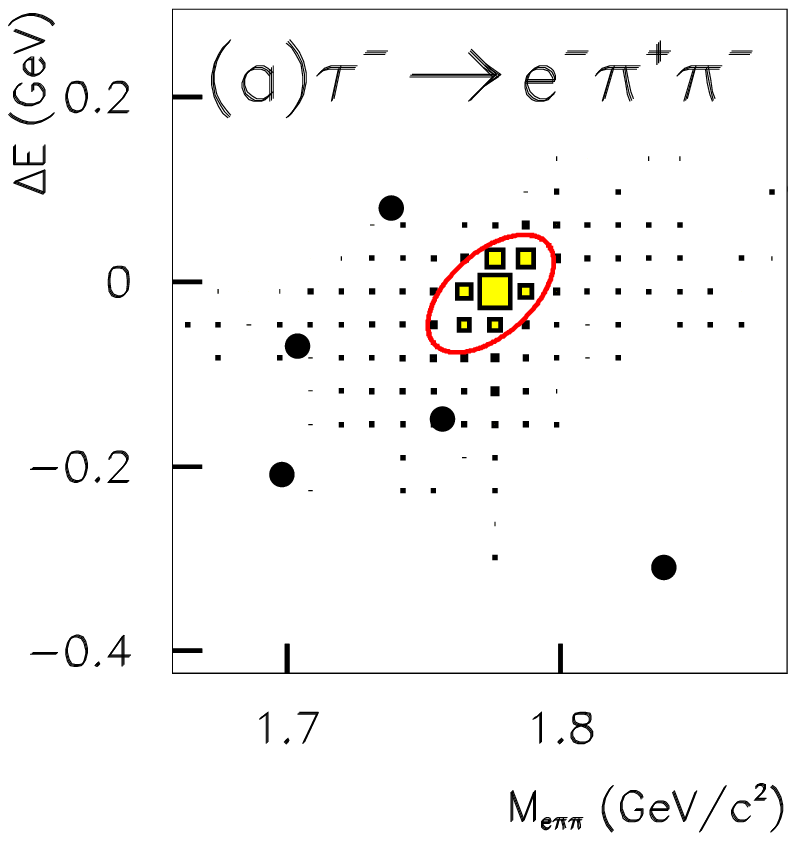}}
 \hspace*{-0.8cm}
       \resizebox{0.35\textwidth}{0.35\textwidth}{\includegraphics
 {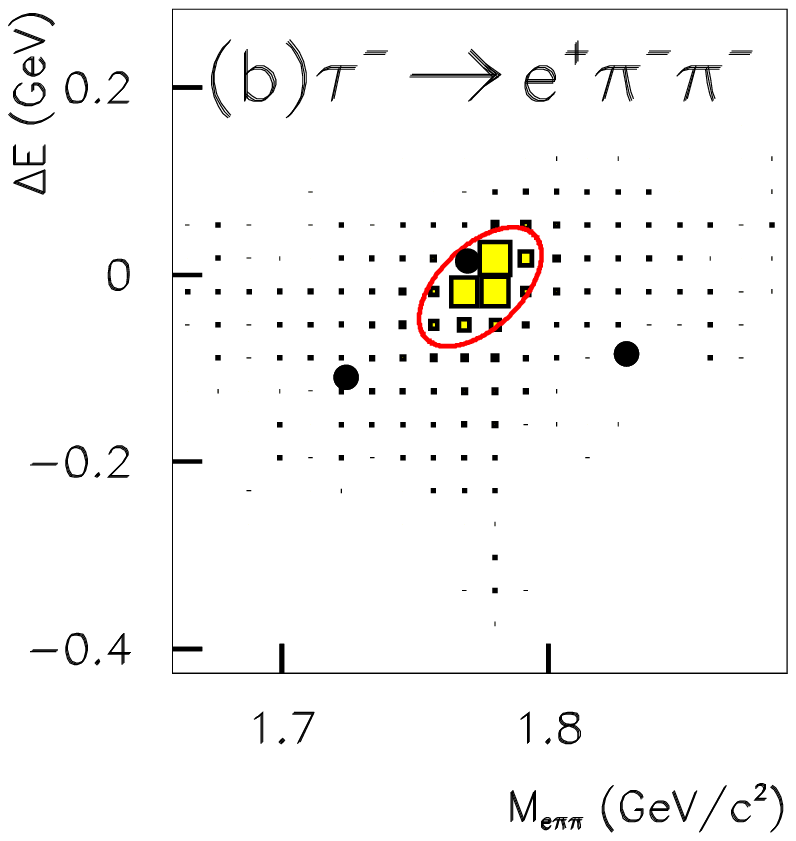}}
\hspace*{-0.8cm}
       \resizebox{0.35\textwidth}{0.35\textwidth}{\includegraphics
 {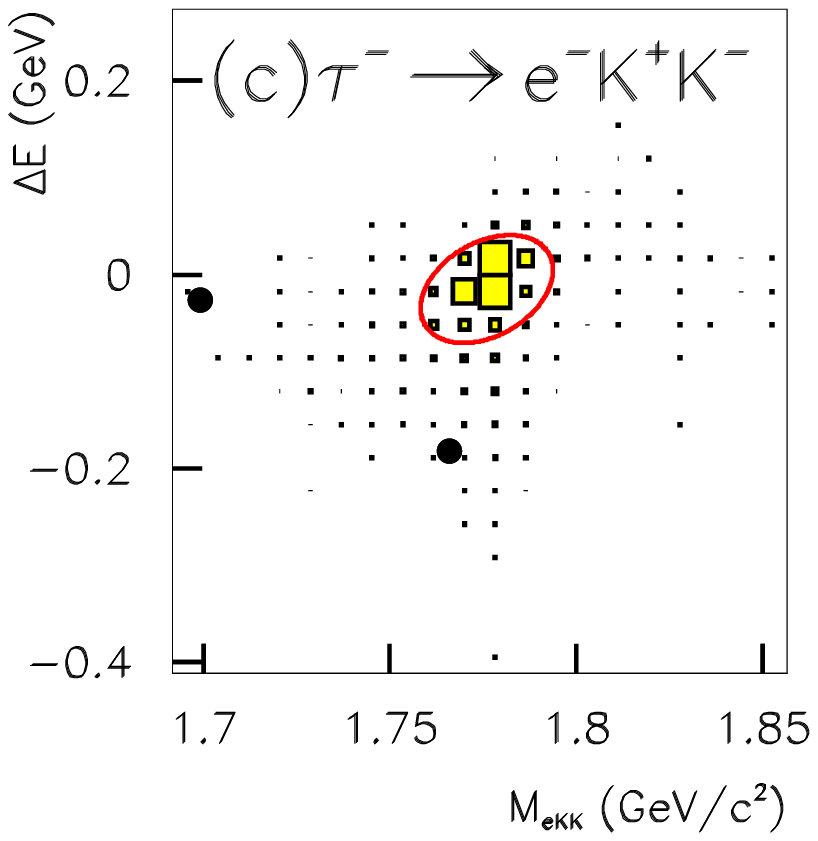}}
       \resizebox{0.35\textwidth}{0.35\textwidth}{\includegraphics
 {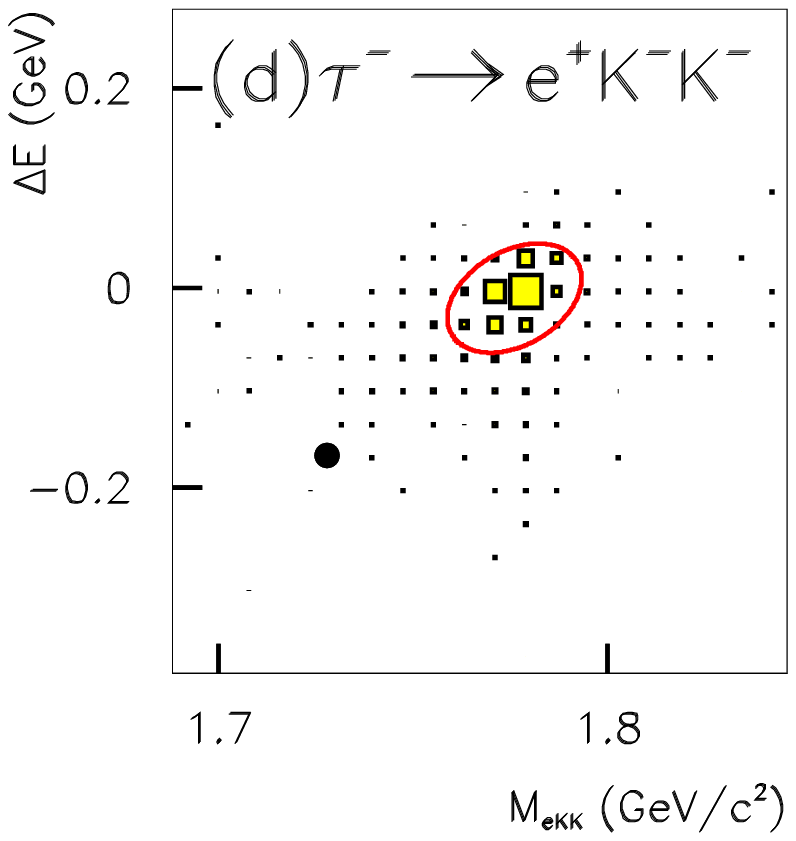}}
\hspace*{-0.8cm}
       \resizebox{0.35\textwidth}{0.35\textwidth}{\includegraphics
 {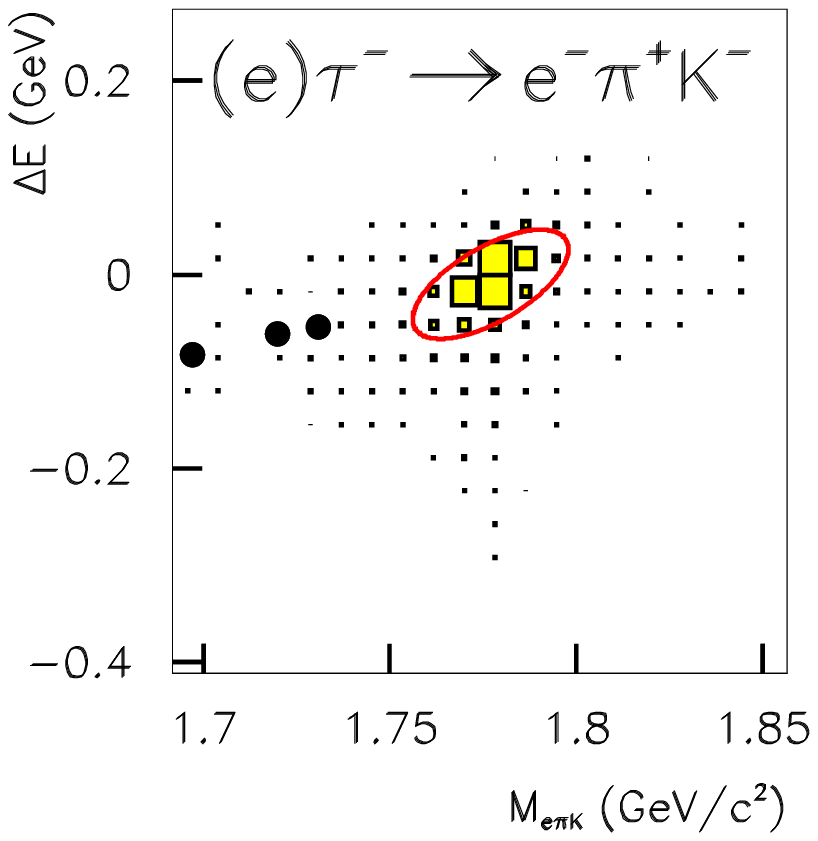}}
\hspace*{-0.8cm}
       \resizebox{0.35\textwidth}{0.35\textwidth}{\includegraphics
 {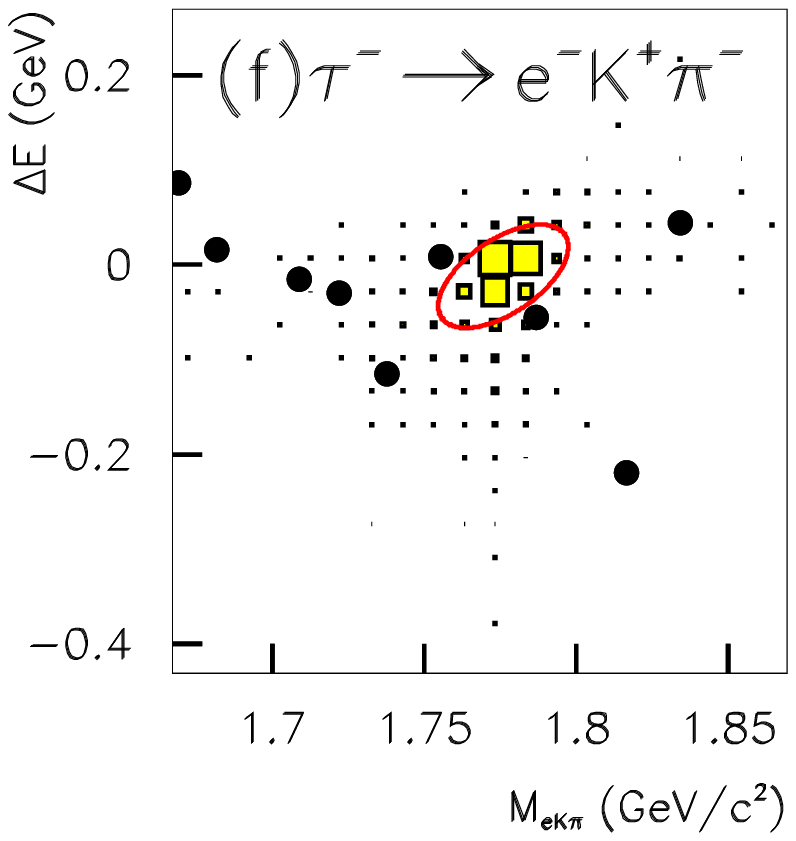}}
       \resizebox{0.35\textwidth}{0.35\textwidth}{\includegraphics
 {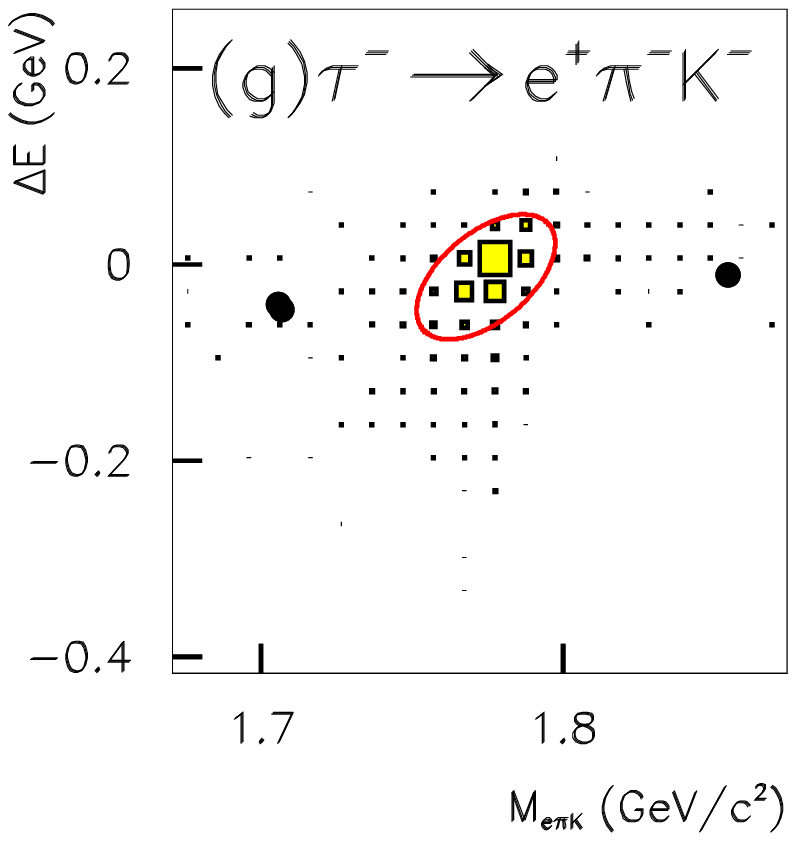}}
 \caption{
Scatter-plots in the $M_{\ell hh'}$ -- $\Delta{E}$ plane  
{within}
the $\pm 20 \sigma$ area for the
(a) $\tau^-\rightarrow e^-\pi^+\pi^-$,
(b) $\tau^-\rightarrow e^+\pi^-\pi^-$,
(c) $\tau^-\rightarrow e^-K^+K^-$,
(d) $\tau^-\rightarrow e^+K^-K^-$,
(e) $\tau^-\rightarrow e^-\pi^+K^-$,
(f) $\tau^-\rightarrow e^-K^+\pi^-$,
and
(g) $\tau^-\rightarrow e^+\pi^-K^-$
modes.
The data are indicated by the solid circles.
The filled boxes show the MC signal distribution
with arbitrary normalization.
The elliptical signal
{regions}
shown by {the} solid {curves}
are used for evaluating the signal yield.
}
\label{fig:ehh}
\end{center}
\end{figure}

\section*{Acknowledgments}

We are grateful to M.J.~Herrero for stimulating discussions.
We thank the KEKB group for the excellent operation of the
accelerator, the KEK cryogenics group for the efficient
operation of the solenoid, and the KEK computer group and
the National Institute of Informatics for valuable computing
and SINET3 network support.  We acknowledge support from
the Ministry of Education, Culture, Sports, Science, and
Technology (MEXT) of Japan, the Japan Society for the 
Promotion of Science (JSPS), and the Tau-Lepton Physics 
Research Center of Nagoya University; 
the Australian Research Council and the Australian 
Department of Industry, Innovation, Science and Research;
the National Natural Science Foundation of China under
contract No.~10575109, 10775142, 10875115 and 10825524; 
the Department of Science and Technology of India; 
the BK21 and WCU program of the Ministry Education Science and
Technology, the CHEP SRC program and Basic Research program (grant No.
R01-2008-000-10477-0) of the Korea Science and Engineering Foundation,
Korea Research Foundation (KRF-2008-313-C00177),
and the Korea Institute of Science and Technology Information;
the Polish Ministry of Science and Higher Education;
the Ministry of Education and Science of the Russian
Federation and the Russian Federal Agency for Atomic Energy;
the Slovenian Research Agency;  the Swiss
National Science Foundation; the National Science Council
and the Ministry of Education of Taiwan; and the U.S.\
Department of Energy.
This work is supported by a Grant-in-Aid from MEXT for 
Science Research in a Priority Area ("New Development of 
Flavor Physics"), and from JSPS for Creative Scientific 
Research ("Evolution of Tau-lepton Physics").

\end{document}